\documentclass
[aps,a4paper,twocolumn,showpacs,showkeys,preprintnumbers,amsmath,amssymb]{revtex4}%
\usepackage{graphicx}
\usepackage{bm}
\usepackage{dcolumn}
\usepackage{color}
\usepackage{pst-plot}
\usepackage{amsmath}
\usepackage{amsfonts}
\usepackage{amssymb}%
\newcommand{\be}{\begin{eqnarray}}
\newcommand{\ee}{\end{eqnarray}}

\begin{document}
\title{A noncommutative approach to the cosmological constant problem}
\author{Remo Garattini}
\thanks{Remo.Garattini@unibg.it}
\affiliation{Facolt\`{a} di Ingegneria, Universit\`{a} degli Studi di Bergamo,  Viale
Marconi 5, 24044 Dalmine (Bergamo) Italy and INFN  Sezione di Milano,
Milan, Italy.}
\author{Piero Nicolini}
\thanks{nicolini@th.physik.uni-frankfurt.de}
\affiliation{Frankfurt Institute for Advanced Studies (FIAS), Institut f\"ur Theoretische
Physik, Johann Wolfgang Goethe-Universit\"at, Ruth-Moufang-Strasse 1, 60438
Frankfurt am Main, Germany}

\begin{abstract}
In this paper we study the cosmological constant emerging from the
Wheeler-DeWitt equation as an eigenvalue of the related Sturm-Liouville
problem. We employ Gaussian trial functionals and we perform a mode
decomposition to extract the transverse-traceless component, namely, the
graviton contribution, at one loop. We implement a noncommutative-geometry-
induced minimal length to calculate the number of graviton modes. As a result,
we find regular graviton fluctuation energies for the Schwarzschild, de Sitter,
and anti-de Sitter backgrounds. No renormalization scheme is necessary to
remove infinities, in contrast to what happens in conventional approaches.
\end{abstract}
\maketitle

%\title{A Non Commutative Approach to the Cosmological Constant }

\section{Introduction}

The emergence of a minimal length is widely accepted as a natural requirement
when quantum features of spacetime are considered. Indeed, the spacetime
structure at small distances is rather different from the conventional
description in terms of a smooth differential manifold. When extreme energies
probe spacetime, quantum gravitational fluctuations appear and prevent any
measure of better accuracy than a natural length scale, e.g., the Planck
length
%, beyond which no further resolution is actually possible
(see, for instance, \cite{Calmet:2004mp}). Qualitatively, we can describe the
spacetime in such an extreme regime as a \textit{quantum foam}, namely, a
complex turbulent storm-tossed sea which accounts for the seething fabric of
the Universe \cite{Wheeler:1957mu}. The presence of a minimal length implies
that singularities in general relativity and ultraviolet divergences in
quantum field theory are nothing but spurious effects due to the inadequacy of
the formalism at small scales/extreme energies, rather than actual physical
phenomena. Along this line of reasoning, the renormalization procedure, too,
even if very effective for its capacity of providing reliable and testable
data, is nothing more than an artificial mechanism to get an \textit{ad hoc}
treatment for the bad short-distance behavior of quantum fields. As a further
criticism to renormalization, there is also the well-known limitation of a
systematic employment of regularization schemes when gravity is taken into
account. A related problem is provided by the calculation of the cosmological
constant:
%%even if it has been clarified that the observed cosmological constant might
%%arise from the fluctuations of the vacuum energy \cite{cosmoenergyfluc},
%%rather than from the vacuum energy itself,
it is not yet clear what is the prescription which leads to a finite and
reasonably small value, since trivial infinity subtractions are not viable in
the presence of a gravitational coupling.
%\cite{Padmanabhan:2004qc}.

Given this background, great efforts have been devoted to implementing a minimal
length in physical theories and curing the aforementioned pathologies or
limitations of conventional approaches. For instance, we recall the route opened
by the generalized uncertainty principle (GUP), according to which the
Heisenberg commutation relation among coordinates and momenta would be
deformed in order to include the effects of an ultraviolet (and/or an infrared)
cut-off \cite{GUP}. In the same spirit, several models of noncommutative
geometry (NCG) have been extensively studied, i.e., geometries for which
coordinate operators might fail to commute, giving rise to an effective
graininess of the spacetime manifold (for general reviews on the topic, see
\cite{NCG}). Even if both the GUP and NCG are often regarded as mere effective
tools or low-energy limits of more fundamental formulations \cite{StringLQGS},
they turn out to be quite successful for their capacity of providing testable
predictions and foreseeing new reliable scenarios \cite{NCpheno}. Among the most
relevant results, we recall that, with a minimal length induced by averaging
noncommutative coordinate fluctuations \cite{NCQFT,NCUnruh,NCspectral}, the
curvature singularity of conventional black hole spacetimes has been tamed
\cite{NCBHs,NCwormhole,NChorizon}, and a new thermodynamically stable final
stage of the Hawking evaporation has been determined \cite{NCthermo,NCstat}
(for a review on these topics, see \cite{review}).
%% As a consequence also the catastrophic behavior
%% of the terminal phase of black hole evaporation has been improved in favor of
%% a thermodynamically stable, positive heat capacity cooling down to a zero
%% temperature black hole remnant configuration \cite{NCthermo,NCwormhole} (for a review on
%% this topics see \cite{review}).
%%
%% Another important result is the derivation of
%% a line element describing a traversable spacetime wormhole, whose throat is
%% sustained and kept open by the anti-gravitational pull consequent of the short
%% distance fluctuating behavior of the quantum manifold \cite{NCwormhole}. If in
%% the above cases, a minimal length has been implemented to model a
%% delocalization of matter which generates the spacetime geometry, recent
%% contributions have been devoted to smear out also matter fields propagating
%% over a given background independently on its curvature. For instance, the
%% propagation of fields with a minimal length has been used to revise the
%% thermal properties of the Unruh-deWitt accelerated detector \cite{NCUnruh} or
%% the stability of spacetimes exhibiting Cauchy horizons \cite{NChorizon}. More
%% recently, the study of a diffusion process has shown in a transparent way,
%% that the spectral dimension of a quantum spacetime is actually two for
%% diffustion time equivalent to the scale governed by the minimal length,
%% suggesting possible renormalization properties for the gravitational
%% interaction \cite{NCspectral}.

In light of the above results, in this paper, we would like to do a step
forward. In particular, we would like to apply some of the NCG properties to
the computation of the cosmological constant. This procedure is based on the
employment of the Wheeler-DeWitt (WDW) equation with the cosmological constant
considered as an eigenvalue of a certain Sturm-Liouville problem. This
approach has been initiated by one of us\cite{remo}, with the purpose of
computing the zero-point energy generated by the graviton fluctuations.
In other words, zero-point energy is a Casimir-like energy. We recall that, for calculating
the Casimir energy, one generally invokes a subtraction procedure between zero-point energies having the same boundary condition. At the semiclassical level, one
employs a zeta function regularization scheme to determine finite energy
densities, when the graviton one-loop contribution to a classical energy is
computed. As a goal of this paper, we want to implement in the WDW equation a
NCG-induced minimal length and show how the resulting zero-point energies
naturally arise as finite quantities without invoking any regularization scheme.

\section{The Wheeler-DeWitt equation and graviton contribution}

\label{p1} The WDW equation is a celebrated equation which formally extends to
the quantum realm the Hamilton-Jacobi equation for general relativity, in the
same fashion of what the Schr\"{o}dinger equation does for quantum mechanics.
It reads
\begin{equation}
\mathcal{H}\Psi=0\label{WDW1}%
\end{equation}
where $\Psi$ is a functional of field configurations on all of spacetime, and
the super-Hamiltonian $\mathcal{H}$ provides a Hamiltonian constraint, i.e.,
restricts $\Psi$ to the physical configuration of the geometry and matter
content of the Universe. The spacetime is supposed to be foliated into a
family of spacelike hypersurfaces $\Sigma$. The Arnowitt-Deser-Misner
 variables offer a valid example of such a foliation.
Explicitly, the metric background is written in the familiar form%
\begin{equation}
ds^{2}=-N^{2}dt^{2}+g_{ij}\left(  N^{i}dt+dx^{i}\right)  \left(
N^{j}dt+dx^{j}\right)  .\label{ds2}%
\end{equation}
$N$ is called the \textit{lapse function} $N$, and $N_{i}$ is the \textit{shift
function}. The dynamical variables are, therefore, the three-dimensional metrics
$g_{ij}(x^{j},t)$, and their conjugate momenta $\pi^{ij}$ are called
supermomenta. The replacement of the dynamical variables with the
corresponding quantum operators
\begin{align}
\hat{g}_{ij}(t,x^{k}) &  \rightarrow g_{ij}(t,x^{k})\\
\hat{\pi}^{ij}(t,x^{k}) &  \rightarrow-i\frac{\delta}{\delta g_{ij}(t,x^{k})}%
\end{align}
provides the quantization. In the following, for brevity, we shall skip the
\textquotedblleft\ $\hat{}$ \textquotedblright\ superscript for operator
notation. In terms of dynamical variables, we can define the
super-Hamiltonian, which reads
\begin{equation}
\mathcal{H}=\left(  2\kappa\right)  G_{ijkl}\pi^{ij}\pi^{kl}-\frac{\sqrt{g}%
}{2\kappa}\left(  \,\!^{3}R-2\Lambda\right)
\end{equation}
where $\kappa=8\pi G$, $G_{ijkl}$ is the supermetric
\[
G_{ijkl}=\frac{1}{2\sqrt{g}}\ \left(  g_{ik}g_{jl}+g_{il}g_{jk}-g_{ij}%
g_{kl}\right)
\]
and $^{3}R$ is the scalar curvature in three dimensions. The main reason to
work with the WDW equation becomes more transparent if we formally rewrite it
as
\begin{align}
&  \frac{1}{V}\ \frac{\int\mathcal{D}\left[  g_{ij}\right]  \ \Psi^{\ast
}\left[  g_{ij}\right]  \ \left(  \ \int_{\Sigma}d^{3}x\ \hat{\Lambda}%
_{\Sigma}\ \right)  \ \Psi\left[  g_{ij}\right]  }{\int\mathcal{D}\left[
g_{ij}\right]  \Psi^{\ast}\left[  g_{ij}\right]  \Psi\left[  g_{ij}\right]
}\label{WDW2}\\
&  =\frac{1}{V}\frac{\left\langle \Psi\left\vert \int_{\Sigma}d^{3}%
x\ \hat{\Lambda}_{\Sigma}\right\vert \Psi\right\rangle }{\left\langle
\Psi|\Psi\right\rangle }=-\frac{\Lambda}{\kappa},\nonumber
\end{align}
where
\begin{equation}
V=\int_{\Sigma}d^{3}x\sqrt{g}%
\end{equation}
is the volume of the hypersurface $\Sigma$, and
\begin{equation}
\hat{\Lambda}_{\Sigma}=\left(  2\kappa\right)  G_{ijkl}\pi^{ij}\pi^{kl}%
-\sqrt{g}^{3}R/\left(  2\kappa\right)  .
\end{equation}
Equation $\left(  \ref{WDW2}\right)  $ represents the Sturm-Liouville problem
associated with the cosmological constant. The related boundary conditions are
dictated by the choice of the trial wave functionals which, in our case, are of
Gaussian type. Different types of wave functionals correspond to different
boundary conditions. We can gain more information if we consider
\[
g_{ij}=\bar{g}_{ij}+h_{ij},
\]
where $\bar{g}_{ij}$ is the background metric and $h_{ij}$ is a quantum
fluctuation around the background. Thus $\left(  \ref{WDW2}\right)  $ can be
expanded in terms of $h_{ij}$. Since the kinetic part of $\hat{\Lambda
}_{\Sigma}$ is quadratic in the momenta, we only need to expand the
three-scalar curvature $\int d^{3}x\sqrt{g}{}^{3}R$ up to the quadratic order.
However, to extract the graviton contribution, we also need an orthogonal
decomposition on the tangent space of three-metric deformations
\cite{Vassilevich}
\begin{equation}
h_{ij}=\frac{1}{3}\left(  \sigma+2\nabla\cdot\xi\right)  g_{ij}+\left(
L\xi\right)  _{ij}+h_{ij}^{\bot}.\label{p21a}%
\end{equation}
The operator $L$ maps the gauge vector $\xi_{i}$ into symmetric tracefree
tensors
\begin{equation}
\left(  L\xi\right)  _{ij}=\nabla_{i}\xi_{j}+\nabla_{j}\xi_{i}-\frac{2}%
{3}g_{ij}\left(  \nabla\cdot\xi\right)  ,
\end{equation}
$h_{ij}^{\bot}$ is the traceless-transverse component of the perturbation
(TT), namely,
\begin{equation}
g^{ij}h_{ij}^{\bot}=0,\qquad\nabla^{i}h_{ij}^{\bot}=0
\end{equation}
and $h$ is the trace of $h_{ij}$. It is immediate to recognize that the trace
element $\sigma=h-2\left(  \nabla\cdot\xi\right)  $ is gauge-invariant. If we
perform the same decomposition also on the momentum $\pi^{ij}$, up to second
order, $\left(  \ref{WDW2}\right)  $ becomes
\begin{equation}
\frac{1}{V}\frac{\left\langle \Psi\left\vert \int_{\Sigma}d^{3}x\left[
\hat{\Lambda}_{\Sigma}^{\bot}+\hat{\Lambda}_{\Sigma}^{\xi}+\hat{\Lambda
}_{\Sigma}^{\sigma}\right]  ^{\left(  2\right)  }\right\vert \Psi\right\rangle
}{\left\langle \Psi|\Psi\right\rangle }=-\frac{\Lambda}{\kappa}%
.\label{lambda0_2}%
\end{equation}
Concerning the measure appearing in $\left(  \ref{WDW2}\right)  $, we have to
note that the decomposition $\left(  \ref{p21a}\right)  $ induces the
following transformation on the functional measure $\mathcal{D}h_{ij}%
\rightarrow\mathcal{D}h_{ij}^{\bot}\mathcal{D}\xi_{i}\mathcal{D}\sigma J_{1}$,
where the Jacobian related to the gauge-vector variable $\xi_{i}$ is%
\begin{equation}
J=\left[  \det\left(  \bigtriangleup g^{ij}+\frac{1}{3}\nabla^{i}\nabla
^{j}-R^{ij}\right)  \right]  ^{\frac{1}{2}}.
\end{equation}
This is nothing but the famous Faddeev-Popov determinant. It becomes more
transparent if $\xi_{a}$ is further decomposed into a transverse part $\xi
_{a}^{T}$, with $\nabla^{a}\xi_{a}^{T}=0$, and a longitudinal part $\xi
_{a}^{\parallel}$, with $\xi_{a}^{\parallel}=$ $\nabla_{a}\psi$. Then, $J$ can
be expressed by an upper triangular matrix for certain backgrounds (e.g.,
Schwarzschild in three dimensions). It is immediate to recognize that, for an
Einstein space in any dimension, cross terms vanish, and $J$ can be expressed
by a block diagonal matrix. Since $\det AB=\det A\det B$, the functional
measure $\mathcal{D}h_{ij}$ factorizes into%
\begin{align}
\mathcal{D}h_{ij} &  =\left(  \det\bigtriangleup_{V}^{T}\right)  ^{\frac{1}%
{2}}\left(  \det\left[  \frac{2}{3}\bigtriangleup^{2}+\nabla_{i}R^{ij}%
\nabla_{j}\right]  \right)  ^{\frac{1}{2}}\nonumber\\
&  \times\mathcal{D}h_{ij}^{\bot}\ \mathcal{D}\xi^{T}\ \mathcal{D}\psi
\end{align}
leading to the Faddeev-Popov determinant with $\left(  \bigtriangleup_{V}%
^{ij}\right)  ^{T}=\bigtriangleup g^{ij}-R^{ij}$ acting on transverse vectors.
In writing the functional measure $\mathcal{D}h_{ij}$, we have here ignored
the appearance of a multiplicative anomaly \cite{EVZ}. Thus, the inner product
can be written as%
\begin{align}
&  \int\mathcal{D}h_{ij}^{\bot}\mathcal{D}\xi^{T}\mathcal{D}\sigma\Psi^{\ast
}\left[  h_{ij}^{\bot}\right]  \Psi^{\ast}\left[  \xi^{T}\right]  \Psi^{\ast
}\left[  \sigma\right]  \Psi\left[  h_{ij}^{\bot}\right]  \Psi\left[  \xi
^{T}\right]  \nonumber\\
&  \times\ \Psi\left[  \sigma\right]  \left(  \det\bigtriangleup_{V}%
^{T}\right)  ^{\frac{1}{2}}\left(  \det\left[  \frac{2}{3}\bigtriangleup
^{2}+\nabla_{i}R^{ij}\nabla_{j}\right]  \right)  ^{\frac{1}{2}}.
\end{align}
Nevertheless, since there is no interaction between ghost fields and the other
components of the perturbation at this level of approximation, the Jacobian
appearing in the numerator and in the denominator simplify. The reason can be
found in terms of connected and disconnected terms. The disconnected terms
appear in the Faddeev-Popov determinant, and the above ones are not linked by the
Gaussian integration. This means that disconnected terms in the numerator and
the same ones appearing in the denominator cancel out. Therefore, $\left(
\ref{lambda0_2}\right)  $ factorizes into three pieces. The piece containing
$E_{\Sigma}^{\bot}$, the contribution of the TT tensors
(TT), is essentially the graviton contribution representing true physical
degrees of freedom. Regarding the vector operator $\hat{\Lambda}_{\Sigma}^{T}%
$, we observe that, under the action of infinitesimal diffeomorphism generated
by a vector field $\epsilon_{i}$, the components of $\left(  \ref{p21a}%
\right)  $ transform as follows \cite{Vassilevich}:%
\begin{equation}
\xi_{j}\longrightarrow\xi_{j}+\epsilon_{j},\qquad h\longrightarrow
h+2\nabla\cdot\xi,\qquad h_{ij}^{\bot}\longrightarrow h_{ij}^{\bot}.
\end{equation}
The Killing vectors satisfying the condition $\nabla_{i}\xi_{j}+\nabla_{j}%
\xi_{i}=0$ do not change $h_{ij}$ and thus should be excluded from the gauge
group. All other diffeomorphisms act on $h_{ij}$ nontrivially. We need to fix
the residual gauge freedom on the vector $\xi_{i}$. The simplest choice is
$\xi_{i}=0.$ This new gauge fixing produces the same Faddeev-Popov determinant
connected to the Jacobian $J$ and, therefore, will not contribute to the final
value. We are left with
\begin{align}
&  \frac{1}{V}\frac{\left\langle \Psi^{\bot}\left\vert \int_{\Sigma}%
d^{3}x\left[  \hat{\Lambda}_{\Sigma}^{\bot}\right]  ^{\left(  2\right)
}\right\vert \Psi^{\bot}\right\rangle }{\left\langle \Psi^{\bot}|\Psi^{\bot
}\right\rangle }+\label{lambda0_2a}\\
&  +\frac{1}{V}\frac{\left\langle \Psi^{\sigma}\left\vert \int_{\Sigma}%
d^{3}x\left[  \hat{\Lambda}_{\Sigma}^{\sigma}\right]  ^{\left(  2\right)
}\right\vert \Psi^{\sigma}\right\rangle }{\left\langle \Psi^{\sigma}%
|\Psi^{\sigma}\right\rangle }=-\frac{\Lambda}{\kappa}.\nonumber
\end{align}
Note that, in the expansion of $\int_{\Sigma}d^{3}x\sqrt{g}{}R$ to second
order, a coupling term between the TT component and the scalar one remains.
However, the Gaussian integration does not allow such a mixing, which has to be
introduced with an appropriate wave functional. By extracting the TT tensor
contribution from $\left(  \ref{WDW2}\right)  $ within second-order
perturbation theory in $h_{ij}$ onto the background $\bar{g}_{ij}$, we get%
\begin{align}
&  [\hat{\Lambda}_{\Sigma}^{\bot}]^{\left(  2\right)  }=\frac{1}{4V}%
\int_{\Sigma}d^{3}x\sqrt{\bar{g}}\ G^{ijkl}\left[  \left(  2\kappa\right)
K^{-1\bot}\left(  x,x\right)  _{ijkl}\right.  \nonumber\\
&  \left.  +\frac{1}{\left(  2\kappa\right)  }\!{}\left(  \tilde
{\bigtriangleup}_{L\!}\right)  _{j}^{a}K^{\bot}\left(  x,x\right)
_{iakl}\right]  ,\label{p22}%
\end{align}
where $\tilde{\bigtriangleup}$ is the modified Lichnerowicz operator
\begin{equation}
\left(  \tilde{\bigtriangleup}_{L\!}\!{}\ h^{\bot}\right)  _{ij}=\left(
\bigtriangleup_{L\!}\!{}\ h^{\bot}\right)  _{ij}-4R{}_{i}^{k}\!{}%
\ h_{kj}^{\bot}+\text{ }^{3}R{}\!{}\ h_{ij}^{\bot}\label{M Lichn}%
\end{equation}
defined in terms of the Lichnerowicz operator
\begin{align}
&  \left(  \bigtriangleup_{L}h\right)  _{ij}=\bigtriangleup h_{ij}%
-2R_{ikjl}h^{kl}+R_{ik}h_{j}^{k}+R_{jk}h_{i}^{k}\nonumber\\
&  \bigtriangleup=-\nabla^{a}\nabla_{a}.\label{DeltaL}%
\end{align}
The metric $G^{ijkl}$ represents the inverse DeWitt supermetric and all
indices run from one to three. Note that the term%
\begin{equation}
-4R{}_{i}^{k}\!{}\ h_{kj}^{\bot}+^{3}R{}\!{}\ h_{ij}^{\bot}%
\end{equation}
disappears in four dimensions when we use a background which is a solution of
the Einstein field equations without matter contribution. The propagator
$K^{\bot}\left(  x,x\right)  _{iakl}$ can be represented as
\begin{equation}
K^{\bot}\left(  \overrightarrow{x},\overrightarrow{y}\right)  _{iakl}%
=\sum_{\tau}\frac{h_{ia}^{\left(  \tau\right)  \bot}\left(  \overrightarrow
{x}\right)  h_{kl}^{\left(  \tau\right)  \bot}\left(  \overrightarrow
{y}\right)  }{2\lambda\left(  \tau\right)  },\label{proptt}%
\end{equation}
where $h_{ia}^{\left(  \tau\right)  \bot}\left(  \overrightarrow{x}\right)  $
are the eigenfunctions of $\tilde{\bigtriangleup}_{L\!}$. The parameter $\tau$
denotes a complete set of indices and $\lambda\left(  \tau\right)  $ are a set
of variational parameters to be determined by the minimization of $\left(
\ref{p22}\right)  $. The expectation value of $\hat{\Lambda}_{\Sigma}^{\bot}$
is easily obtained by inserting the form of the propagator into $\left(
\ref{p22}\right)  $ and minimizing with respect to the variational function
$\lambda\left(  \tau\right)  $. As a result, the expectation value of
$\hat{\Lambda}_{\Sigma}^{\bot}$ can be written in terms of the eigenvalues
$\omega_{i}^{2}\left(  \tau\right)  $ of $\tilde{\bigtriangleup}_{L\!}$. By
means of $\left(  \ref{lambda0_2a}\right)  $, we obtain a cosmological term due
to the TT tensor one-loop energy density%
\begin{equation}
\frac{\Lambda}{8\pi G}=-\frac{1}{2V}\sum_{\tau}\left[  \sqrt{\omega_{1}%
^{2}\left(  \tau\right)  }+\sqrt{\omega_{2}^{2}\left(  \tau\right)  }\right]
,\label{1loop}%
\end{equation}
provided $\omega_{i}^{2}\left(  \tau\right)  >0$. The above expression is
interpreted as the expectation value of graviton fluctuations on a given
background. In the above calculation, we did not consider the scalar
contribution coming from $\Lambda_{\Sigma}^{\sigma}$, since, in the physically
relevant cases, it is possible to show that it does not contribute. To complete
the picture, we need to specify the form of the background $\bar{g}_{ij}$. In
the next section, we will work within the spherically symmetric case.

\section{The spherically symmetric background}

The line element $\left(  \ref{ds2}\right)  $ can be recast in the following
form:
\begin{equation}
ds^{2}=-N^{2}\left(  r\right)  dt^{2}+\frac{dr^{2}}{1-\frac{b\left(  r\right)
}{r}}+r^{2}\left(  d\theta^{2}+\sin^{2}\theta d\phi^{2}\right)
,\label{metric}%
\end{equation}
where $b\left(  r\right)  $ is termed the \textquotedblleft\textit{shape
function.}\textquotedblright With the help of the Regge and Wheeler
representation,
%\cite{Regge Wheeler},
$\left(  \tilde{\bigtriangleup}_{L\!}\!{}h^{\bot}\right)  _{ij}$ reduces to%
\begin{align}
&  \left[  -\frac{d^{2}}{dx^{2}}+\frac{l\left(  l+1\right)  }{r^{2}}+m_{i}%
^{2}\left(  r\right)  \right]  f_{i}\left(  x\right)  =\omega_{i,l}^{2}%
f_{i}\left(  x\right)  \nonumber\\
&  i=1,2\qquad\qquad\left(  r\equiv r\left(  x\right)  \right)  ,\label{p34}%
\end{align}
where we have used reduced fields of the form $f_{i}\left(  x\right)
=F_{i}\left(  x\right)  /r$ and where we have defined two r-dependent
effective masses $m_{1}^{2}\left(  r\right)  $ and $m_{2}^{2}\left(  r\right)
$:%
\[
\left\{
\begin{array}
[c]{c}%
m_{1}^{2}\left(  r\right)  =\frac{6}{r^{2}}\left(  1-\frac{b\left(  r\right)
}{r}\right)  +\frac{3}{2r^{2}}b^{\prime}\left(  r\right)  -\frac{3}{2r^{3}%
}b\left(  r\right)  \\
\\
m_{2}^{2}\left(  r\right)  =\frac{6}{r^{2}}\left(  1-\frac{b\left(  r\right)
}{r}\right)  +\frac{1}{2r^{2}}b^{\prime}\left(  r\right)  +\frac{3}{2r^{3}%
}b\left(  r\right)
\end{array}
\right.
\]
with $x$ as the proper distance from the throat at $r_{t}=b(r_{t})$, i.e.,
\[
dx=\pm\frac{dr}{\sqrt{1-\frac{b(r)}{r}}}.
\]
There are two interesting cases where a symmetry in the masses appears. The
first case is the Schwarzschild metric with $r_{t}=b\left(  r_{t}\right)
=2MG$. Thus, masses $m_{1}^{2}\left(  r\right)  $ and $m_{2}^{2}\left(
r\right)  $ read%
\begin{equation}
\left\{
\begin{array}
[c]{c}%
m_{1}^{2}\left(  r\right)  =\frac{6}{r^{2}}\left(  1-\frac{2MG}{r}\right)
-\frac{3MG}{r^{3}}\\
\\
m_{2}^{2}\left(  r\right)  =\frac{6}{r^{2}}\left(  1-\frac{2MG}{r}\right)
+\frac{3MG}{r^{3}}.
\end{array}
\right.
\end{equation}
In the range where $r\in\left[  2MG,5MG\right]  $, we have
\begin{equation}
m_{1}^{2}\left(  r\right)  =-m_{2}^{2}\left(  r\right)  =m_{0}^{2}\left(
r\right)  .\label{emS}%
\end{equation}
The second case comes from the de Sitter (dS) [anti-de Sitter (AdS)] metric
with $b\left(  r\right)  =\frac{\Lambda_{dS}}{3}r^{3}$ $\left(  -\frac
{\Lambda_{AdS}}{3}r^{3}\right)  $. Thus, $m_{1}^{2}\left(  r\right)  $ and
$m_{2}^{2}\left(  r\right)  $ become
\[
\left\{
\begin{array}
[c]{c}%
m_{1}^{2}=m_{2}^{2}=m_{dS}^{2}=\frac{6}{r^{2}}\left(  1-\frac{\Lambda_{dS}}%
{3}r^{2}\right)  +\Lambda_{dS}\\
\\
m_{1}^{2}=m_{2}^{2}=m_{AdS}^{2}=\frac{6}{r^{2}}\left(  1+\frac{\Lambda_{AdS}%
}{3}r^{2}\right)  -\Lambda_{AdS}.
\end{array}
\right.
\]
Note that in the case of the dS background, $r\in\left[  0,\sqrt{3/\Lambda
_{dS}}\right]  $, while, for the AdS background, one works in the range
$r\in\left[  0,\infty\right)  $. In order to use the WKB approximation along
the lines of the `t Hooft brick wall problem \cite{'tHooft:1984re}, we can
extract two r-dependent radial wave numbers from $\left(  \ref{p34}\right)$:
\begin{equation}
k_{i}^{2}\left(  r,l,\omega_{i,nl}\right)  =\omega_{i,nl}^{2}-\frac{l\left(
l+1\right)  }{r^{2}}-m_{i}^{2}\left(  r\right)  \quad i=1,2\quad.\label{kTT}%
\end{equation}
It is now possible to explicitly evaluate $\left(  \ref{1loop}\right)  $ in
terms of the effective masses. To further proceed, we have to count the number
of modes with frequency less than $\omega_{i}$, $i=1,2$. This is given
approximately by%
\begin{equation}
\tilde{g}\left(  \omega_{i}\right)  =\int_{0}^{l_{\max}}\nu_{i}\left(
l,\omega_{i}\right)  \left(  2l+1\right)  dl,\label{p41}%
\end{equation}
where $\nu_{i}\left(  l,\omega_{i}\right)  $, $i=1,2$ is the number of nodes
in the mode with $\left(  l,\omega_{i}\right)  $, such that
\begin{equation}
\nu_{i}\left(  l,\omega_{i}\right)  =\frac{1}{\pi}\int_{-\infty}^{+\infty
}dx\sqrt{k_{i}^{2}\left(  r,l,\omega_{i}\right)  }.\label{p42}%
\end{equation}
Here it is understood that the integration with respect to $x$ and $l$ is
taken over those values which satisfy $k_{i}^{2}\left(  r,l,\omega_{i}\right)
\geq0,$ $i=1,2$. However, $\left(  \ref{p41}\right)  $ is based on the
classical Liouville counting number of nodes%
\begin{equation}
dn=\frac{d^{3}\vec{x}d^{3}\vec{k}}{\left(  2\pi\right)  ^{3}}.
\end{equation}
The procedure leads to divergent results. Conventionally, one performs a
renormalization absorbing the divergent parts into the redefinition of bare
classical quantities. In the spirit of any efficient quantum gravity approach,
such a procedure must be reviewed. Indeed, both GUP and NCG formulations
predict a deformation of the integration measure in momentum space,
\begin{equation}
1=\int\frac{d^{n}k}{\left(  1+\mathcal{F}(\vec{k}^{2})\right)  }\left\vert
k\right\rangle \left\langle k\right\vert .
\end{equation}
The function $\mathcal{F}(\vec{k}^{2})$ depends on positive powers of the
argument. As a result $\mathcal{F}(\vec{k}^{2})$, accounts for the suppression
in the UV region, when an effective minimal length models the quantum gravity
uncertainty. As shown in \cite{NCthermo,NCstat}, NCG in coherent-state
formalism provides a specific form for the function $\mathcal{F}(\vec{k}^{2}%
)$. Thus, the number of states reads
%% while, in our case, we are interested in modification coming from the
%% non-commutative theory. With the help of Eq.$\left(  \ref{kTT}\right)  $ and
%% following Ref.\cite{ZH}, we find that%
\begin{equation}
dn=\frac{d^{3}\vec{x}d^{3}\vec{k}}{\left(  2\pi\right)  ^{3}}\ \Longrightarrow
\ dn_{i}=\frac{d^{3}\vec{x}d^{3}\vec{k}}{\left(  2\pi\right)  ^{3}}\exp\left(
-\frac{\theta}{4}k_{i}^{2}\right)  ,\label{moddn}%
\end{equation}
with%
\begin{equation}
k_{i}^{2}=\omega_{i,nl}^{2}-m_{i}^{2}\left(  r\right)  \quad i=1,2.
\end{equation}
This deformation corresponds to an effective cut-off on the background
geometry (\ref{metric}). The UV cut-off is triggered only by higher-momenta
modes $\gtrsim1/\sqrt{\theta}$ which propagate over the background geometry.
The virtue of this kind of deformation lies in the fact that the exponential
damping not only fulfils the general requirement of UV completeness for fields
$f_{i}(x)$, but also provides the strongest possible suppression of higher
momenta. Even if we are dealing with an effective approach that, strictly
speaking, can reliably work only until scales $\sim\sqrt{\theta}$, this
exponential profile lets us have at least a glimpse at smaller scales. To this
purpose, we recall that this kind of deformation of the integration measure has
been already successfully employed in taming the nonperturbative behavior of
the gravitational field: curvature singularities in general relativity have
been cured, giving rise to new quantum corrected regular geometries also at
black hole centers without any breakdown at small scales \cite{NCBHs}.
Plugging $\left(  \ref{p42}\right)  $ into $\left(  \ref{p41}\right)  $ and
taking account of $\left(  \ref{moddn}\right)  $, the number of modes with
frequency less than $\omega_{i}$, $i=1,2$ is given by%
\begin{align}
&  \tilde{g}\left(  \omega_{i}\right)  =\frac{1}{\pi}\int_{-\infty}^{+\infty
}dx\int_{0}^{l_{\max}}\sqrt{\omega_{i,nl}^{2}-\frac{l\left(  l+1\right)
}{r^{2}}-m_{i}^{2}\left(  r\right)  }\nonumber\\
&  \times\left(  2l+1\right)  \exp\left(  -\frac{\theta}{4}k_{i}^{2}\right)
\ dl.
\end{align}
After integration over modes, one gets
\begin{align}
&  \tilde{g}\left(  \omega_{i}\right)  =\frac{2}{3\pi}\int_{-\infty}^{+\infty
}dx\ r^{2}\left[  \frac{3}{2}\sqrt{\left(  \omega_{i,nl}^{2}-m_{i}^{2}\left(
r\right)  \right)  ^{3}}\right.  \nonumber\\
&  \left.  \exp\left(  -\frac{\theta}{4}\left(  \omega_{i,nl}^{2}-m_{i}%
^{2}\left(  r\right)  \right)  \right)  \right]  .
\end{align}
This form of $\tilde{g}\left(  \omega_{i}\right)  $ allows an integration by
parts in $\left(  \ref{1loop}\right)$, leading to%
\begin{align}
\frac{\Lambda}{8\pi G} &  =-\frac{1}{4\pi^{2}V}\sum_{i=1}^{2}\int_{0}%
^{+\infty}\omega_{i}\frac{d\tilde{g}\left(  \omega_{i}\right)  }{d\omega_{i}%
}d\omega_{i}\nonumber\\
&  =\frac{1}{4\pi^{2}V}\sum_{i=1}^{2}\int_{0}^{+\infty}\tilde{g}\left(
\omega_{i}\right)  d\omega_{i}.\label{t1loop}%
\end{align}
This is the graviton contribution to the induced cosmological constant at one
loop. To get this result, we have used $\left(  \ref{emS}\right)  $ and we have
included an additional $4\pi$ coming from the angular integration. As a result
for the Schwarzschild case, we find for the energy%
\begin{equation}
4\pi\int_{-\infty}^{+\infty}dx\ r^{2}\left[  \frac{\Lambda}{8\pi G}-\frac
{1}{4\pi^{2}}\sum_{i=1}^{2}\int_{0}^{+\infty}\tilde{g}\left(  \omega
_{i}\right)  d\omega_{i}\right]  =0.
\end{equation}
Extracting the energy density we find%
\begin{align}
\frac{\Lambda}{8\pi G} &  =\frac{1}{6\pi^{2}}\left[  \int_{\sqrt{m_{0}%
^{2}\left(  r\right)  }}^{+\infty}\sqrt{\left(  \omega^{2}-m_{0}^{2}\left(
r\right)  \right)  ^{3}}e^{-\frac{\theta}{4}\left(  \omega^{2}-m_{0}%
^{2}\left(  r\right)  \right)  }\right.  \nonumber\\
&  +\left.  \int_{0}^{+\infty}\sqrt{\left(  \omega^{2}+m_{0}^{2}\left(
r\right)  \right)  ^{3}}e^{-\frac{\theta}{4}\left(  \omega^{2}+m_{0}%
^{2}\left(  r\right)  \right)  }\right]  .\label{Schw1loop}%
\end{align}
In the Appendix \ref{app}, we explicitly evaluate the previous integrals. Plugging
the result of $\left(  \ref{I1I2}\right)  $ into $\left(  \ref{Schw1loop}%
\right)  $, we get%
\begin{align}
&  \frac{\Lambda}{8\pi G}=\frac{1}{12\pi^{2}}\left(  \frac{4}{\theta}\right)
^{2}\left(  y\cosh\left(  \frac{y}{2}\right)  -y^{2}\sinh\left(  \frac{y}%
{2}\right)  \right)  \nonumber\\
&  \times\ K_{1}\left(  \frac{y}{2}\right)  +y^{2}\cosh\left(  \frac{y}%
{2}\right)  K_{0}\left(  \frac{y}{2}\right)  ,\label{LambdaNCS}%
\end{align}
where%
\begin{equation}
y=\frac{m_{0}^{2}\left(  r\right)  \theta}{4}=\frac{3MG\theta}{4r^{3}%
}.\label{xS}%
\end{equation}
The asymptotic properties of $\left(  \ref{LambdaNCS}\right)  $ show that the
one-loop contribution is  regular everywhere. Indeed, when we rescale the
radial coordinate to the wormhole throat
\[
\rho\equiv\frac{r}{2MG}%
\]
with $\rho\in\lbrack1,5/2]$, we have
\begin{equation}
y=\frac{1}{8\rho^{3}}\frac{\theta}{(MG)^{2}}.
\end{equation}
This means that, when $MG\ll\theta$, we have $y\rightarrow\infty$. From the
expression $\left(  \ref{asym}\right)  $, we find that, when $y\rightarrow
+\infty$,
\begin{align}
\frac{\Lambda}{8\pi G} &  \simeq\frac{1}{12\pi^{2}}\left(  \frac{4}{\theta
}\right)  ^{2}\\
&  \times\left\{  \frac{1}{8}\sqrt{\frac{\pi}{y}}\left[  3+\left(
8y^{2}+6y+3\right)  \exp\left(  -y\right)  \right]  \right\}  \rightarrow
0\nonumber\label{LNCSz}%
\end{align}
%% $\Lambda\to 0$,
%% go to zero,
%% we find%
%% \begin{equation}
%% x=\frac{m_{0}^{2}\left(  r\right)  \theta}{4}=\frac{3MG\theta}{4r^{3}}%
%% =\frac{3MG\theta}{4\left(  \rho2MG\right)  ^{3}}%
%% \end{equation}
%% with $\rho=r/2MG$ and%
%% \begin{equation}
%% x\rightarrow\infty\qquad\mathrm{when\qquad}M\rightarrow0
%% \end{equation}
%% leading to a vanishing cosmological constant,
namely, we recover the correct behavior, according to which, for a vanishing
background gravity, i.e., $M=0$, the one-loop energy must go to zero.
Conversely, when $MG\gg\theta$, we have $y\rightarrow0$ and, from expression
$\left(  \ref{seri}\right)  $, we obtain%
\begin{align}
\frac{\Lambda}{8\pi G} &  \simeq\frac{1}{12\pi^{2}}\left(  \frac{4}{\theta
}\right)  ^{2}\\
&  \times\left[  2-\left(  \frac{7}{8}+\frac{3}{4}\ln\left(  \frac{y}%
{4}\right)  +\frac{3}{4}\gamma\right)  y^{2}\right]  \rightarrow\frac{8}%
{3\pi^{2}\theta^{2}}.\nonumber
\end{align}
a finite value for $\Lambda$. This shows the effect of the NCG cut-off
$\sqrt{\theta}$ at work.

%% The asymptotic properties of Eq.$\left(  \ref{LambdaNCS}\right)  $ show that
%% the one loop contribution is everywhere regular. Indeed, when we rescale the
%% radial coordinate to the wormhole throat
%% \be
%% \rho\equiv \frac{r}{2MG}
%% \ee
%% with $\rho\in [1, 5/2]$ we have
%% \begin{equation}
%% x= \frac{1}{8\rho^3} \frac{\theta}{(MG)^2}.
%% \end{equation}
%% This means that when $MG\ll \theta$, we have $\Lambda\to 0$,
%% go to zero,
%% we find%
%% \begin{equation}
%% x=\frac{m_{0}^{2}\left(  r\right)  \theta}{4}=\frac{3MG\theta}{4r^{3}}%
%% =\frac{3MG\theta}{4\left(  \rho2MG\right)  ^{3}}%
%% \end{equation}
%% with $\rho=r/2MG$ and%
%% \begin{equation}
%% x\rightarrow\infty\qquad\mathrm{when\qquad}M\rightarrow0
%% \end{equation}
%% leading to a vanishing cosmological constant,
%% namely we recover the correct
%% behavior according to which for a vanishing background gravity, i.e. $M=0$, the one loop energy must go to %% zero.
For the dS and AdS cases, we find that the effective masses contribute in the
same way at one loop. Thus, $\left(  \ref{t1loop}\right)  $ becomes
\begin{equation}
\frac{\Lambda}{8\pi G}=2\times\frac{1}{6\pi^{2}}\int_{\sqrt{m_{0}^{2}\left(
r\right)  }}^{+\infty}\sqrt{\left(  \omega^{2}-m_{0}^{2}\left(  r\right)
\right)  ^{3}}e^{-\frac{\theta}{4}\left(  \omega^{2}-m_{0}^{2}\left(
r\right)  \right)  }.
\end{equation}
Plugging the result of $\left(  \ref{I1}\right)  $ into $\left(
\ref{t1loop}\right)  $, we get%
\begin{align}
\frac{\Lambda}{8\pi G}  &  =\frac{1}{6\pi^{2}}\left(  \frac{4}{\theta}\right)
^{2}\\
&  \times\left(  \frac{1}{2}z\left(  1-z\right)  K_{1}\left(  \frac{z}%
{2}\right)  +\frac{1}{2}z^{2}K_{0}\left(  \frac{z}{2}\right)  \right)
\exp\left(  \frac{z}{2}\right)  , \nonumber\label{LNCdSAdS}%
\end{align}
where%
\begin{equation}
\left\{
\begin{array}
[c]{c}%
z=m_{dS}^{2}\left(  r\right)  \theta/4\\
\mathrm{or}\\
z=m_{AdS}^{2}\left(  r\right)  \theta/4.
\end{array}
\right.
\end{equation}
To analyze these results, we recall that, in the de Sitter case, the radial
coordinates $r\in\left[  0,\sqrt{3/\Lambda_{dS}}\right]  $. Therefore, at short
distances $r\ll\sqrt{\theta}$, we have
\[
z=\frac{3}{2}\frac{\theta}{r^{2}}-\frac{\Lambda_{dS}\theta}{4}\rightarrow
\infty.
\]
From expansions $\left(  \ref{asy}\right)  $ and $\left(  \ref{ser}\right)$,
we find%
\begin{equation}
\frac{\Lambda}{8\pi G}\simeq\frac{1}{6\pi^{2}}\left(  \frac{4}{\theta}\right)
^{2}\frac{3}{8}\sqrt{\frac{\pi}{z}}\rightarrow0,
\end{equation}
when $z\rightarrow\infty$. This corresponds to the correct behavior in a
spacetime region where the curvature vanishes. On the other hand, for
$r\approx\sqrt{3/\Lambda_{dS}}\gg\sqrt{\theta}$, we have
\[
z\approx\frac{\Lambda_{dS}\theta}{4}\rightarrow0
\]
which implies
\begin{align}
&  \frac{\Lambda}{8\pi G}\simeq\frac{1}{6\pi^{2}}\left(  \frac{4}{\theta
}\right)  ^{2}\\
&  \times\left[  1-\frac{z}{2}+\left(  -\frac{7}{16}-\frac{3}{8}\ln\left(
\frac{z}{4}\right)  -\frac{3}{8}\gamma\right)  z^{2}\right]  \rightarrow
\frac{8}{3\pi^{2}\theta^{2}},\nonumber
\end{align}
i.e., a finite value of the cosmological term. The same conclusion holds for
the anti-de Sitter case.
%%, while when $x\rightarrow0$, one gets%
%% The asymptotic properties of Eq.$\left(  \ref{LambdaNCS}\right)  $ show that
%% the one loop contribution is everywhere regular. Indeed, when we rescale the
%% radial coordinate to the wormhole throat and we let the mass $M$ go to zero,
%% we find%
%% \begin{equation}
%% x=\frac{m_{0}^{2}\left(  r\right)  \theta}{4}=\frac{3MG\theta}{4r^{3}}%
%% =\frac{3MG\theta}{4\left(  \rho2MG\right)  ^{3}}%
%% \end{equation}
%% with $\rho=r/2MG$ and%
%% \begin{equation}
%% x\rightarrow\infty\qquad\mathrm{when\qquad}M\rightarrow0
%% \end{equation}
%% leading to a vanishing cosmological constant, namely we recover the correct
%% behavior that for a vanishing background the one loop energy must be
%% vanishing. The same behavior is obtained for the dS and AdS geometries.
%%
%% %\begin{center}
%% %{\LARGE Sembra che non sia necessario introdurre il BH diffuso!!}
%% %\end{center}
%%
%% This is not surprising, because the shape function can be approximated by%
%% \begin{equation}
%% b(r)\simeq\frac{2r_{t}}{\sqrt{\pi}}\gamma\left(  \frac{3}{2},\frac{r^{2}%
%% }{4\theta}\right)  \simeq\frac{r_{t}r^{3}}{6\sqrt{\pi\theta^{3}}}%
%% =\frac{\Lambda_{NC}r^{3}}{3},
%% \end{equation}
%% representing a de Sitter metric in the range $r^{2}/4\theta\ll1$.
%%

\section{Summary and Conclusions}

\label{p6}

\begin{figure}[h]
\centering
\includegraphics[width=2.8in]{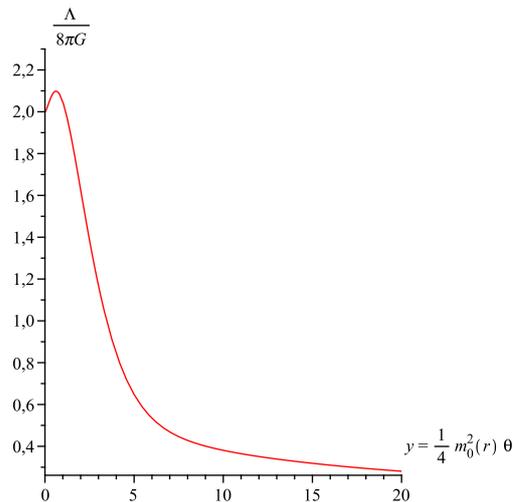}\caption{Plot of $\Lambda/8\pi G$ in
Planck units as a function of the scale-invariant $y$, which depends on the
background choice. For dS and AdS backgrounds, the variable $y$ is replaced by
$z.$}%
\label{Lambda}%
\end{figure}

In this paper, we calculated the cosmological constant as an eigenvalue of the
Sturm-Liouville problem related to the Wheeler-DeWitt equation. With the help
of Gaussian trial functionals, we extracted the one-loop contribution of the
transverse-traceless component, namely, the graviton. Instead of embarking in
conventional regularization schemes, we implemented a natural UV cut-off in
the background geometry, invoking a NCG-induced minimal length. As a result, we
get a modified counting of graviton modes. This lets us obtain 
regular values everywhere for the cosmological constant, independently of the chosen
background, which, nevertheless, is of a spherically symmetric type. We show
this for the Schwarzschild, de Sitter and anti-de Sitter backgrounds. The
strength of our approach lies in the specific kind of integration measure
deformation in momentum space we derived from NCG. This lets us overcome
previous attempts which only led to mild effects and just a reduction of the
degree of divergence \cite{statecounting,NCCCP}. Although the result seems to
be promising, we have to note that the evaluation is at the Planck scale, and,
even if Fig. $\left(  \ref{Lambda}\right)  $ shows a vanishing behavior,
one has to bear in mind that this behavior corresponds to the switching off of
the Schwarzschild background. The paper is subjected to future developments.
First, we restricted the attention only on spherically symmetric backgrounds
like Schwarzschild or de Sitter/anti-de Sitter backgrounds. A further extension
should be the inclusion of rotations, which considerably increase the
technical difficulty level. Moreover, regarding the Schwarzschild background,
we worked with the \textquotedblleft\textit{classical Schwarzschild}%
\textquotedblright\ and not with the smeared solution predicted by the
noncommutative theory developed in configuration space having a shape function
$b\left(  r\right)  $ of the form%
\begin{equation}
b_{NC}(r)=\frac{4MG}{\sqrt{\pi}}\gamma\left(  \frac{3}{2},\frac{r^{2}}%
{4\theta}\right)  \,.
\end{equation}
The use of $b_{NC}(r)$ instead of $b\left(  r\right)  $ could introduce new
features of the full noncommutative theory, allowing a better exploration of
the wormhole throat. As a further point, we have to observe that, even if we
have a finite value for the cosmological constant, it will still come too
large with respect to its observed value. This seems to be a general fact, as
far as one employs a UV natural cut-off \cite{Padmanabhan:2004qc}. A possible
solution to this problem could be found in the fact that the cosmological
constant might arise from fluctuations of vacuum energy \cite{cosmoenergyfluc}%
, rather than from the vacuum energy itself. Therefore, we believe that the
paper is opening a new route to further investigations. 

\begin{acknowledgments}
\noindent P.N. is supported by the Helmholtz International Center for FAIR
within the framework of the LOEWE Program (Landesoffensive zur Entwicklung
Wissenschaftlich-\"{O}konomischer Exzellenz) launched by the State of Hesse.
\end{acknowledgments}

\appendix

\section{Integrals}

\label{app} In this Appendix, we explicitly compute the integrals coming from
$\left(  \ref{t1loop}\right)  $. We begin with
%\begin{widetext}%
\begin{align}
&  \int_{\sqrt{m_{0}^{2}\left(  r\right)  }}^{+\infty}\sqrt{\left(  \omega
^{2}-m_{0}^{2}\left(  r\right)  \right)  ^{3}}e^{-\frac{\theta}{4}\left(
\omega^{2}-m_{0}^{2}\left(  r\right)  \right)  }d\omega\label{I1a}\\
&  \underset{\omega^{2}=x}{=}\frac{1}{2}\int_{\sqrt{m_{0}^{2}\left(  r\right)
}}^{+\infty}\sqrt{\left(  x-m_{0}^{2}\left(  r\right)  \right)  ^{3}}%
e^{-\frac{\theta}{4}\left(  x-m_{0}^{2}\left(  r\right)  \right)  }\frac
{dx}{\sqrt{x}}\nonumber\\
&  =\exp\left(  \frac{m_{0}^{2}\left(  r\right)  \theta}{4}\right)  \frac
{1}{2}\left(  \frac{\theta}{4}\right)  ^{-\frac{3}{2}}\sqrt{m_{0}^{2}\left(
r\right)  }\Gamma\left(  \frac{5}{2}\right) \nonumber\\
&  \times\exp\left(  -\frac{m_{0}^{2}\left(  r\right)  \theta}{8}\right)
W_{-1,-1}\left(  \frac{m_{0}^{2}\left(  r\right)  \theta}{4}\right)
,\nonumber
\end{align}
%\end{widetext}
where we have used the following relationship%
\begin{align}
&  \int_{u}^{+\infty}x^{\nu-1}\left(  x-u\right)  ^{\mu-1}e^{-\beta x}dx=\\
&  \beta^{-\frac{\nu+\mu}{2}}u^{\frac{\nu+\mu-2}{2}}\Gamma\left(  \mu\right)
\exp\left(  -\frac{\beta u}{2}\right)  W_{\frac{\nu-\mu}{2},\frac{1-\nu-\mu
}{2}}\left(  \beta u\right) \nonumber\\
&  \operatorname{Re}\mu>0\quad\operatorname{Re}\beta u>0,\nonumber
\end{align}
where $W_{\mu,\nu}\left(  x\right)  $ is the Whittaker function and
$\Gamma\left(  \nu\right)  $ is the gamma function. Further manipulation on
$\left(  \ref{I1a}\right)  $ leads to%
\begin{align}
&  \frac{1}{2}\left(  \frac{\theta}{4}\right)  ^{-2}\left(  \frac{1}%
{2}x\left(  1-x\right)  K_{1}\left(  \frac{x}{2}\right)  +\frac{1}{2}%
x^{2}K_{0}\left(  \frac{x}{2}\right)  \right) \nonumber\\
&  \times\exp\left(  \frac{x}{2}\right)  , \label{I1}%
\end{align}
where%
\begin{equation}
x=\frac{m_{0}^{2}\left(  r\right)  \theta}{4}.
\end{equation}
It is useful to write an asymptotic expansion for $K_{0}\left(  \frac{x}%
{2}\right)  $ and $K_{1}\left(  \frac{x}{2}\right)  $. We get
\begin{equation}%
\begin{array}
[c]{c}%
K_{0}\left(  x/2\right)  \simeq\sqrt{\pi}e^{-x/2}x^{-\frac{1}{2}}\left(
1-\frac{1}{4x}\right)  +O\left(  {x}^{-\frac{5}{2}}\right) \\
K_{1}\left(  x/2\right)  \simeq\sqrt{\pi}e^{-x/2}x^{-\frac{1}{2}}\left(
1+\frac{3}{4x}\right)  +O\left(  {x}^{-\frac{5}{2}}\right)
\end{array}
. \label{Knu}%
\end{equation}
Plugging expansion $\left(  \ref{Knu}\right)  $ into expression $\left(
\ref{I1}\right)  $, one obtains that the asymptotic behavior is given by%
\begin{align}
& \frac{1}{2}\left(  \frac{\theta}{4}\right)  ^{-2}\times\\
& \sqrt{\pi}\left(  \frac{1}{2}\sqrt{x}\left(  1-x\right)  \left(  1+\frac
{3}{4x}\right)  +\frac{1}{2}\sqrt{x^{3}}\left(  1-\frac{1}{4x}\right)  \right)
\nonumber\\
&  +O\left(  {x}^{-\frac{5}{2}}\right) \nonumber
\end{align}
and after a further simplification, one gets%
\begin{equation}
\frac{1}{2}\left(  \frac{\theta}{4}\right)  ^{-2}\frac{3}{8}\sqrt{\frac{\pi
}{x}} \label{asy}%
\end{equation}
while when $x\rightarrow0$, one gets%
\begin{equation}
\frac{1}{2}\left(  \frac{\theta}{4}\right)  ^{-2}\left[  1-\frac{x}{2}+\left(
-\frac{7}{16}-\frac{3}{8}\ln\left(  \frac{x}{4}\right)  -\frac{3}{8}%
\gamma\right)  x^{2}\right]  . \label{ser}%
\end{equation}
For the other integral, we proceed in the same way and we get
%\begin{widetext}%
\begin{align}
&  \int_{0}^{+\infty}\sqrt{\left(  \omega^{2}+m_{0}^{2}\left(  r\right)
\right)  ^{3}}e^{-\frac{\theta}{4}\left(  \omega^{2}+m_{0}^{2}\left(
r\right)  \right)  }d\omega\label{I2a}\\
&  =\exp\left(  -\frac{m_{0}^{2}\left(  r\right)  \theta}{8}\right)  \frac
{1}{2}\left(  \frac{\theta}{4}\right)  ^{-\frac{3}{2}}\sqrt{m_{0}^{2}\left(
r\right)  }\nonumber\\
&  \times\ \Gamma\left(  \frac{1}{2}\right)  W_{1,-1}\left(  \frac{m_{0}%
^{2}\left(  r\right)  \theta}{4}\right)  .\nonumber
\end{align}
%\end{widetext}
Converting to Bessel functions, $\left(  \ref{I2a}\right)  $ yields%
\begin{equation}
\frac{1}{2}\left(  \frac{\theta}{4}\right)  ^{-2}\left(  \frac{x}{2}\left(
1+x\right)  K_{1}\left(  \frac{x}{2}\right)  +\frac{x^{2}}{2}K_{0}\left(
\frac{x}{2}\right)  \right)  \exp\left(  -\frac{x}{2}\right)  , \label{I2}%
\end{equation}
whose sum with Eq.$\left(  \ref{I1}\right)  $ gives%
\begin{align}
&  \frac{1}{2}\left(  \frac{\theta}{4}\right)  ^{-2}\left(  x\cosh\left(
\frac{x}{2}\right)  -x^{2}\sinh\left(  \frac{x}{2}\right)  \right)
K_{1}\left(  \frac{x}{2}\right) \nonumber\\
&  +x^{2}\cosh\left(  \frac{x}{2}\right)  K_{0}\left(  \frac{x}{2}\right)  .
\label{I1I2}%
\end{align}
Thus the asymptotic expansion for $\left(  \ref{I1I2}\right)  $ yields%
\begin{equation}
\frac{1}{2}\left(  \frac{\theta}{4}\right)  ^{-2}\left\{  \frac{1}{8}%
\sqrt{\frac{\pi}{x}}\left[  3+\left(  8x^{2}+6x+3\right)  \exp\left(
-x\right)  \right]  \right\}  . \label{asym}%
\end{equation}
On the other hand, when $x\rightarrow0$, one gets%
\begin{equation}
\frac{1}{2}\left(  \frac{\theta}{4}\right)  ^{-2}\left[  2-\left(  \frac{7}%
{8}+\frac{3}{4}\ln\left(  \frac{x}{4}\right)  +\frac{3}{4}\gamma\right)
x^{2}\right]  . \label{seri}%
\end{equation}

\end{document}